\documentclass[10pt]{article}

\usepackage{amsmath}
\usepackage{amssymb}

\usepackage{graphicx}

\usepackage{cite}

\usepackage{color} 

\usepackage{setspace} 

\usepackage{fullpage}

\usepackage[labelfont=bf,labelsep=period,justification=raggedright]{caption}

\makeatletter
\renewcommand{\@biblabel}[1]{\quad#1.}
\makeatother

\date{\today}

\newcommand{\Var}{\rm Var}

\newcommand{\E}{\textrm{E}}

\newcommand{\Na}{\textrm{Na}}
\newcommand{\K}{\textrm{K}}
\newcommand{\Nap}{$\textrm{Na}^+$}
\newcommand{\Kp}{$\textrm{K}^+$}
\newcommand{\vect}[1]{\mathbf{#1}}

\newcommand{\beq}{\begin{equation}}
\newcommand{\eeq}{\end{equation}}
\newcommand{\beqr}{\begin{eqnarray}}
\newcommand{\eeqr}{\end{eqnarray}}
\newcommand{\beqrn}{\begin{eqnarray*}}
\newcommand{\eeqrn}{\end{eqnarray*}}
\newcommand{\beqn}{\begin{equation*}}
\newcommand{\eeqn}{\end{equation*}}
\newcommand{\bei}{\begin{itemize}}
\newcommand{\eei}{\end{itemize}}
\newcommand{\ii}{\begin{itemize} \item}
\newcommand{\beii}{\begin{itemize} \item}

\newcommand{\bes}{\begin{small}}
\newcommand{\ees}{\end{small}}
\newcommand{\bec}{\begin{center}}
\newcommand{\eec}{\end{center}}

\begin{document}

\title{The what and where of adding channel noise to the Hodgkin-Huxley equations}


\author{Joshua H. Goldwyn$^{1}$, 
Eric Shea-Brown$^{1,2}$
\\
$^1$ Department of Applied Mathematics, University of Washington, Seattle, WA, USA
\\
$^2$ Program in Neurobiology and Behavior, University of Washington, Seattle, WA, USA
\\
E-mail: jgoldwyn@uw.edu
}
\maketitle

\section*{Abstract} 
One of the most celebrated successes in computational biology is the Hodgkin-Huxley framework for modeling electrically active cells.  This framework, expressed through a set of differential equations, synthesizes the impact of ionic currents on a cell's voltage --  and the highly nonlinear impact of that voltage back on the currents themselves -- into the rapid push and pull of the action potential.  Latter studies confirmed that these cellular dynamics are orchestrated by individual ion channels, whose conformational changes regulate the conductance of each ionic current.  Thus, kinetic equations familiar from physical chemistry are the natural setting for describing conductances; for small-to-moderate numbers of channels, these will predict fluctuations in conductances and stochasticity in the resulting action potentials.  At first glance, the kinetic equations provide a far more complex (and higher-dimensional) description than the original Hodgkin-Huxley equations.  This has prompted more than a decade of efforts to capture channel fluctuations with noise terms added to the Hodgkin-Huxley equations.  Many of these approaches, while intuitively appealing, produce quantitative errors when compared to kinetic equations; others, as only very recently demonstrated, are both accurate and relatively simple.  We review what works, what doesn't, and why, seeking to build a bridge to well-established results for the deterministic Hodgkin-Huxley equations.  As such, we hope that this review will speed emerging studies of how channel noise modulates electrophysiological dynamics and function.  We supply user-friendly Matlab simulation code of these stochastic versions of the Hodgkin-Huxley equations on the ModelDB website (accession number 138950) and http://www.amath.washington.edu/$\sim$etsb/tutorials.html.


\section*{Introduction}

Understanding the role of noise in cellular dynamics and function is a central challenge across computational biology.  
This is as true in neuroscience as in any field~\cite{Faisal2008, Laing2010, Rolls2010}, and a universal source of noise in electrically active cells that has garnered increasing attention is the stochastic activity in ion channels~\cite{Sakmann1995,White2000,Hille2001}.   This {\it channel noise}  has been studied in a variety of neural systems including electrical stimulation of the auditory nerve by cochlear implants~\cite[e.g.]{Imennov2009, Woo2010}, as well as in entorhinal cortex~\cite{White1998}, cerebellar granule cells~\cite{Saarinen2008}, and hippocampal CA1 pyramidal neurons~\cite{Cannon2010}.  Modeling studies have suggested that channel noise can influence information processing~\cite{Sengupta2010}, spike time reliability~\cite{Schneidman1998}, stochastic resonance~\cite{Schmid2001}, firing irregularity~\cite{Rowat2007, Saarinen2008}, subthreshold dynamics~\cite{White1998, Saarinen2008}, and action potential initiation and propagation in morphologically detailed models~\cite{Faisal2007, Cannon2010}.  Channel noise is at work in many other systems such as the activity of cold receptor cells~\cite{Finke2008}, nicotinic acetylcholine receptors~\cite{Keleshian2000}, and calcium release by Inositol 1,4,5-Triphosphate receptors~\cite{Shuai2002}.
 
Despite widespread interest in channel noise, it has remained unclear what the options are for including this noise source in the canonical model of neurophysiology --  the Hodgkin-Huxley (HH) equations for the action potential~\cite{Hodgkin1952}.  The direct approach provides a gold standard:  each of $N$ channels of a particular type transitions independently and randomly among discrete configurational states.  This yields a continuous-time Markov chain with voltage-dependent transition probabilities; see~\cite{Groff2010} for a recent review.  In the limit that $N \rightarrow \infty$ for each channel type, the classical Hodgkin-Huxley equations are recovered~\cite{Fox1994, Dayan2001, Austin2008, Keener2009, Pakdaman2010, Goldwyn2011}.  For finite $N$, one simulates the Markov process via a Gillespie-type algorithm~\cite{Gillespie1977, Skaugen1979, Chow1996, Faisal2007}.  

Is there a simpler approach, where one modifies the HH equations by adding a few well-placed noise terms?  Beyond conceptual and computational simplicity, this would provide a direct link to powerful results on the dynamics and geometry of these equations~\cite{Izhikevich2007, RinzelErmentrout1998}.  This line of research was initiated by Fox and Lu~\cite{Fox1994, Fox1997} who derived candidate sets of stochastic differential equations (SDEs) using a system-size expansion applied to the Markov chain.   The past few years have seen increasing interest in this problem, spurred on by the promise yet apparent shortcomings of this SDE approach~\cite{Mino2002, Zeng2004, Bruce2007a, Saarinen2008, Bruce2009, Goldwyn2011, Linaro2011, Orio2011}.  

As recent work attests~\cite{Goldwyn2011, Linaro2011, Orio2011}, accurate methods for incorporating channel noise into the HH equations are finally emerging in the form of methods both new and old.  These works demonstrate that adding noise terms to the HH equations can indeed give a compressed and accurate reproduction of the channel fluctuations.
However, the placement of these terms is critical, and -- as a decade of research attests --  less than obvious.  A key focus of our review is a unified presentation of the methods that provide the most accurate approximations to Markov chain models of channel noise.
A common feature of these methods is that they introduce noise processes as \emph{conductances} in the HH equations.

\section*{Stochastic versions of the Hodgkin-Huxley Equations}

As they are an almost universal point of reference for neuron modeling, we consider the equations introduced by Hodgkin and Huxley to model action potentials in the squid giant axon~\cite{Hodgkin1952}. 
\begin{align}
C \frac{dV}{dt} &= -\bar{g}_\Na m^3 h(V-E_\Na) - \bar{g}_\K n^4(V-E_\K) -g_\textrm{L} (V-E_\textrm{L}) + I    \label{eq:V_HH} \\
\frac{dx}{dt} &= \alpha_x (1-x) - \beta_x x  \label{eq:x_HH} , \mbox{ where } x=m,h, \mbox{ or } n.
\end{align}
Here, $V$ is the membrane voltage, the gating variables $m$, $h$, and $n$ represent the fraction of open channel \emph{subunits} of different types, aggregated across the entire cell membrane.  These fractions are combined in the terms $m^3 h$ and $n^4$ to regulate total conductances for \Nap and \Kp currents.  The constant $C$ represents the capacitance of the cell membrane; $E_\Na$, $E_\K$, and $E_\textrm{L}$ are reversal potentials; $\bar{g}_\Na$ and $\bar{g}_\K$ are maximal conductances; and $g_\textrm{L}$ is the leak conductance.  

Comprehensive introductions to this model can be found in many standard texts~\cite{Dayan2001, Izhikevich2007, RinzelErmentrout1998}.  We emphasize that our discussion applies to any conductance-based model of excitable cells, including point, compartmental, or spatially extended neurons, and related models of calcium release \cite{Li1994}.  To model channel noise within this differential equation framework, we seek ways of introducing fluctuations into this (classically) deterministic system.  In this review, we consider three approaches, which we classify as follows:

 \begin{itemize}
 \item \textbf{Current noise}: Replace Eq.~\ref{eq:V_HH} with
 	\begin{align}
	 \tag{\ref{eq:V_HH}*}
 	C \frac{dV}{dt} &= -\bar{g}_\Na m^3 h(V-E_\Na) - \bar{g}_\K n^4(V-E_\K) -g_\textrm{L} (V-E_\textrm{L}) + I  + \xi_V(t)  \label{eq:V_HH_CurrentNoise}
 	\end{align}
	where $\xi_V(t)$ is a Gaussian white noise process.  
 \item \textbf{Subunit noise}: Replace Eq.~\ref{eq:x_HH} with
	\begin{align}
	 \tag{\ref{eq:x_HH}*}
 	\frac{dx}{dt} &= \alpha_x (1-x) - \beta_x x  +\xi_x(t) \label{eq:x_HH_SubunitNoise}, \mbox{ where } x=m,h, \mbox{ or } n.
	\end{align}
	where the $\xi_x(t)$ are Gaussian processes that may depend on $x$ and $V$. 
 \item \textbf{Conductance noise}   Replace Eq.~\ref{eq:V_HH} with
  	\begin{align}
	 \tag{\ref{eq:V_HH}**}
	C \frac{dV}{dt} &= -\bar{g}_\Na (m^3 h + \xi_\Na) (V-E_\Na(t)) - \bar{g}_\K (n^4+\xi_\K(t))(V-E_\K) -g_\textrm{L} (V-E_\textrm{L}) + I  \label{eq:V_HH_ConductanceNoise}
	 \end{align}
	 where the noise processes $\xi_\Na(t)$ and $\xi_\K(t)$ are Gaussian processes that may depend on $x$ and $V$.
\end{itemize}

Table 1 summaries the differences among these models, which we now discuss in detail.

 \begin{table}[!ht]
\caption{
\bf{Classification of channel noise models}}
\begin{tabular}{|c|c|c|c|c|}
Noise Model & Voltage dynamics & Subunit dynamics & Fraction open \Nap channels & Fraction open \Kp channels \\
None & Eq.~\ref{eq:V_HH}& Eq.~\ref{eq:x_HH} & $m^3 h$ &  $n^4$ \\
Current & Eq.~\ref{eq:V_HH_CurrentNoise}& Eq.~\ref{eq:x_HH} & $m^3 h$ &  $n^4$ \\
Subunit & Eq.~\ref{eq:V_HH}& Eq.~\ref{eq:x_HH_SubunitNoise} & $m^3 h$ &  $n^4$ \\
Conductance & Eq.~\ref{eq:V_HH_ConductanceNoise}& Eq.~\ref{eq:x_HH} & $m^3 h + \xi_\Na(t)$ &  $n^4+ \xi_\K(t)$
\end{tabular}
\begin{flushleft} A summary of the three classes of channel noise models that we discuss in this review, and how they differ from the deterministic HH equations which have no noise.
\end{flushleft}
\label{tab:label}
\end{table}

\subsection*{Current noise}
 \label{subsec:CurrentNoise}
The simplest method for incorporating noise into the HH equations is to add a fluctuating current term  $\xi_V(t)$ to the $dV/dt$ equation, as shown in Eq.~\ref{eq:V_HH_CurrentNoise}.   Here, we assume $\xi_V(t)$ is only a function of time.  Stochastic currents of this form are frequently used to drive the HH model, often in the context of the diffusion approximation for synaptic inputs~\cite{Gerstein1964, Tuckwell1988, Brunel2001}.  In the present context, however, we emphasize that $\xi_V(t)$ is meant to represent the combined effect of the stochastic activity of ion channels on the voltage dynamics of the cell.  This approach is appealing due to its simplicity, but since channel noise is generated by the stochastic activity of ion channels in the cell membrane, it seems likely that the fluctuation term $\xi_V(t)$ should also depend on $V$ or the subunit variables.  Another drawback is that, to date, there is no principled method for determining the intensity of the noise.  Nonetheless, there may be cases in which current noise can be justified on empirical grounds.   For instance, for a single membrane area and a constant applied current, Rowat compared the interspike interval distribution generated by a Markov chain model to the distribution generated by HH equations with current noise and found remarkably close agreement~\cite{Rowat2007}.

\subsection*{Subunit noise}
\label{subsec:SubunitNoise}

In the HH framework, an ion channel's configuration is determined by the states of its constituent subunits, where each subunit can be either in an open or closed state~\cite{Johnston1995, Dayan2001, Hille2001}.   For instance, each \Kp channel is composed of four $n$-type subunits, all of which must be open in order for the channel to be permeable to \Kp ions.
Each subunit randomly transitions between its open and closed state.  This suggests that the most appropriate place to add noise may be to the equations that describe the fractions of open \emph{subunits}, as in Eq.~\ref{eq:x_HH_SubunitNoise}.  Moreover, since all subunits are independent and all subunits of the same type are statistically identical, it is tempting to combine the resulting noisy fractions of open subunits to regulate conductances in the same way as one would in the deterministic HH equations; namely by computing $m^3 h$ and $n^4$.

The variables $m$, $h$, and $n$ represent the aggregated fraction of open \emph{subunits}, whereas the quantity that influences the membrane potential is the fraction of individual open \emph{channels}.  In the limit of infinitely many channels (and therefore vanishing fluctuation terms), $m^3 h$ and $n^4$  do correctly model the fraction of open channels.  For a finite number of channels, however, there is no guarantee that fluctuations in the these quantities will correctly model fluctuations in the membrane-wide fractions of open channels.     

To see this, note that if all channels were gated by a single subunit, then the subunit model would be appropriate --- in this case, the (noisy) fraction of open subunits is identical to the (noisy) fraction of open channels.  In the HH model, however, four subunits gate each channel.  
Combining the quantities $m$, $h$, and $n$ together to form the quantities $m^3 h$ and $n^4$ neglects the important fact that each ion channel is composed of a specific package of subunits.  The states of the particular subunits within a channel, not the average state of all subunits in the cell membrane, determine whether that channel is open or closed.  Thus, random transitions of individual channels among their different configurational states occur with different statistics than predicted by random transitions of the aggregated subunit variables alone~\cite{Goldwyn2011}.  This fact leads to quantitative errors produced by the subunit noise approach, as we will review below.

Subunit noise was first proposed in~\cite{Fox1994} and has been used many times; see ~\cite{DeVries2000, Schmid2001, Shuai2002, Casado2003, Rowat2004, Wang2004, Jo2005, Ozer2005, Saarinen2008, Finke2008, Cudmore2010, Sato2010}, among others.
By applying a system size expansion to the states of populations of subunits, Fox and Lu arrived at a Langevin equation description of the subunit dynamics, precisely of the form of Eq.~\ref{eq:x_HH_SubunitNoise}, where the noise terms $\xi_x(V,t)$ ($x=m,h, \mbox{ or } n$) are Gaussian processes with covariance function
\begin{align}
\label{eq:FoxLuSubunitCovar}
\E [\xi_x(t), \xi_x(t')] = \frac{\alpha_x (1-x) + \beta_x x}{N} \delta(t-t').
\end{align}
Here, $\delta(\cdot)$ is the Dirac delta function and $N$ represents either the number of \Nap channels for the $m$ and $h$ subunits or the number of \Kp channels for the $n$ subunit.  Although the authors acknowledged that the subunit noise approach has no rigorous justification and must be validated empirically, it has been widely used as an approximation to Markov chain ion channel models.  However, numerical studies have revealed inaccuracies in this approximation that persist even as the number of channels increases~\cite{Zeng2004, Bruce2009}.  Relative to the Markov chain model, the subunit noise models produce weaker conductance and voltage fluctuations~\cite{Bruce2009, Faisal2010}, lower firing rates~\cite{Sengupta2010} (and, equivalently, longer mean interspike intervals~\cite{Zeng2004}), less variability in the occurrences and timing of spikes in response to a brief pulse of current~\cite{Mino2002, Bruce2007a}, and they transmit information at a higher rate~\cite{Sengupta2010}. Furthermore, mathematical analyses of the voltage clamp statistics of the subunit noise model have proven that it does not generate stationary distributions of open channels that accurately approximate those of the Markov chain model~\cite{Goldwyn2011, Linaro2011}.  

The analysis in~\cite{Goldwyn2011} revealed similar inaccuracies in a related model proposed by~\cite{Shuai2002}, in which the terms $m^3 h$ and $n^4$ terms in Eq.~\ref{eq:V_HH} are replaced by $m_1 m_2 m_3 h$ and $n_1 n_2 n_3 n_4$, respectively, where the subscript denotes independent solutions to SDEs of the form of Eq~\ref{eq:x_HH_SubunitNoise}.  Others have proposed simplifying Eq.~\ref{eq:FoxLuSubunitCovar} so that the noise terms do not depend on $V$, and are simply Gaussian white noise~\cite{Saarinen2008}.  While such approaches may be justified on empirical grounds, in general they should not be considered as systematic approximations to Markov chain ion channel models.

\subsection*{Conductance noise}
\label{subsec:ConductanceNoise}

The remaining possibility is to incorporate fluctuations directly into the fractions of open channels.  This seems natural, as the fraction of open channels controls ionic currents.  Our intuitive understanding of the HH equations, which can be made rigorous as in~\cite{Dayan2001, Keener2009, Goldwyn2011}, tells us that the mean fractions of open \Nap and \Kp channels are given by $m^3 h$ and $n^4$.  The most direct approach to adding channel noise to the HH equations, therefore, is to add zero mean stochastic processes to the deterministic values of $m^3 h$ and $n^4$.  Following this idea leads to Eq.~\ref{eq:V_HH_ConductanceNoise}, which is a compact mathematical description of channel noise that preserves the original structure of the HH equations and has the biophysically desirable interpretation that channel noise induces fluctuations in the ionic conductances.  We now review three channel noise models~\cite{Fox1994, Goldwyn2011, Linaro2011} and, with a brief set of calculations, place them in the unified framework  of conductance noise.

\subsubsection*{Conductance noise models based on voltage clamp}
\label{subsubsec:VoltageClamp}

Two recent studies have developed conductance noise models based on stationary statistics of channel activity in voltage clamp --- called the ``quasistationary" channel model in~\cite{Goldwyn2011} and the ``effective" model in~\cite{Linaro2011}.   Using the standard assumption that all ion channels are independent, the stationary distribution of open channels in voltage clamp is a binomial distribution parameterized by the total number of channels and the probability that any given channel is open.  The probability that a channel is open depends on $V$, and thus a voltage clamp analysis generates a family of binomial distributions indexed by $V$, which is treated as a fixed parameter.  The means of the distributions of open channels are given by familiar terms from the deterministic HH equations: $m^3 h$ for \Nap channels and $n^4$ for \Kp channels. If these binomial distributions are well approximated by Gaussian distributions, then the stationary distribution of open channels in voltage clamp can be accurately approximated by a family of zero mean, voltage dependent Gaussian processes that are added to the voltage dependent equilibrium values of $m^3 h$ and $n^4$.  

The effective model of~\cite{Linaro2011}, for instance, represents the fraction of open \Kp channels in voltage clamp as $n^4 + \xi_\K(V,t)$ where the stochastic processes  are sums of independent Ornstein-Uhlenbeck (OU) processes (i.e. Gaussian colored noise).  In other words, $\xi_\K(V,t) = \sum_i \zeta_i(V,t)$, where the $\zeta_i(t)$ are defined by SDEs of the form:
\begin{align}
\label{eq:OU}
d \zeta_i(V,t) = -\frac{\zeta_i(V,t)}{\tau_{i}(V)}dt + \sigma_{i}(V) dW_{i}(t)
\end{align}
with timescales $\tau_{i}(V)$ and noise amplitudes $\sigma_{i}(V)$~\cite{Goldwyn2011,Linaro2011}.  The quasistationary channel model in~\cite{Goldwyn2011} produces equivalent Gaussian processes in voltage clamp.  The difference between the two methods is that, in~\cite{Goldwyn2011}, there is a single noise process shared by all OU processes: $dW_i(t) = dW(t)$ for all $i$ in Eq.~\ref{eq:OU}.  While this leads to different values of $\sigma_i(V)$, our own simulations of these models (not shown) did not reveal any systematic differences in the outputs of the two models. 

To simulate these models with a freely evolving membrane potential, one must assume Eq.~\ref{eq:OU} is valid outside of voltage clamp.  In practice, one numerically integrates Eq.~\ref{eq:OU}, where $V$ is updated in each time step according to Eq.~\ref{eq:V_HH_ConductanceNoise}.  There is no assurance that this approach is valid in the context of a dynamic membrane potential.  If $V$ changes on longer time scales than the correlation times in the conductance fluctuations, then such an approximation may be appropriate, but a defining feature of neural dynamics is the rapid change in $V$ during the course of an action potential.   Voltage clamp-based methods may be less reliable, therefore, for modeling the spiking activity of neurons.

These channel noise models were developed in~\cite{Goldwyn2011} and~\cite{Linaro2011} in order to approximate the original Markov chain description of channel kinetics.  Their structural details --  i.e. the number of $\zeta_i(V,t)$ processes used to define $\xi_K(V,t)$ and $\xi_{Na}(V,t)$ and the values of $\tau_i(V)$ and $\sigma_i(V)$ in Eq.~\ref{eq:OU} -- were defined based on the stationary statistics of Markov chain model.  The voltage clamp approach itself, however, can be made general and model-independent.  The only necessary ingredients are the autocovariance functions, as a function of the voltage clamp value, for fluctuations in the conductances.  Moreover, if these stationary autocovariance functions can be expressed as sums of exponential functions, then the Gaussian representation theory for multiple Markov processes ensures that they can be approximated as a linear combination of  OU processes~\cite{Hida1993}.

\subsubsection*{Conductance noise models based on Fox and Lu's system size expansion}
\label{subsub:SystemSize}

Lacking in all of the previously discussed methods is a direct approach for modeling the dynamics of fluctuations in the fractions of open channels as the voltage $V$ dynamically evolves.  Surprisingly, the early work of Fox and Lu addressed this problem, but has apparently gone overlooked ever since.  Fox and Lu derived a system of SDEs in which each dynamical variable represents the fraction of ion channels in a specified configuration.  This differs from their more widely-used model, the subunit model discussed previously, in which the dynamical variables represent the fractions of open subunits.  The resulting system of SDEs does not visibly resemble the HH equations, but with a few calculations we next show that this approach produces a conductance noise model in the form of Eq.~\ref{eq:V_HH_ConductanceNoise}. 

The starting point of Fox and Lu's analysis are vectors that describe the fractions of \Nap and \Kp channels in each configuration as a function of time.  We denote these by $\vect{y}(t)$ and $\vect{x}(t)$.  For instance, the elements of $\vect{x}$ represent the fraction of \Kp channels that have all subunits closed, three subunits closed and one open, etc.  The state that will be of most interest is the conducting state, in which all subunits are open.  We denote the corresponding elements of $\vect{y}$ and $\vect{x}$  as $y_o$ and $x_o$, and write the current balance equation as:
\begin{align}
\label{eq:V_OpenVar}
C \frac{dV}{dt} &= -\bar{g}_\Na y_o (V-E_\Na) - \bar{g}_\K x_o (V-E_\K) -g_\textrm{L} (V-E_\textrm{L}) + I 
\end{align}

The dynamics of $y_o$ and $x_o$ are determined by drift and diffusion matrices, which Fox and Lu obtained from the original Markov chain description through a system size expansion~\cite{Fox1994, Fox1997,Gardiner2004}.   We omit the details of the system size expansion, which can be found in~\cite{Fox1994, Fox1997}.  A rigorous discussion of a related method for passing from the Markov chain kinetics to a system of SDEs has been recently presented~\cite{Pakdaman2010},  and we do not reproduce those details here.  
The result of Fox and Lu's expansion is a coupled system of linear SDEs of the form:
\begin{align}
d \vect{y} &= A_\Na(V)\vect{y} dt + S_\Na(V,\vect{y}) d\vect{W_\Na}(t). \label{eq:FL12D_Na} \\
d \vect{x} &= A_\K(V)\vect{x} dt + S_\K(V,\vect{x}) d\vect{W_\K}(t) \label{eq:FL12D_K} .
\end{align}
The matrices $A_\Na(V)$ and $A_\K(V)$ are the drift term or deterministic part of the dynamics, and are identical to the transition matrices from the master equation representation of the Markov chains for the \Nap and \Kp channels~\cite{Fox1994, Keener2009, Goldwyn2011}.  The matrices $S_\Na(V,\vect{y})$ and $S_\K(V,\vect{x})$ are matrix square roots of diffusion matrices; they depend on the state variable and the voltage-dependent transition rates.  Stochasticity arises  via the independent, standard Brownian 
processes $\mathbf{W}_\Na(t)$ and  $\mathbf{W_\K}(t)$.

We now demystify the connection between these equations, in which fractions of open channels are obtained from a high-dimensional system of coupled SDEs, and the standard HH equations, in which the fractions of open channels depend on the subunit variables.  The key is to split the equations for $\vect{x}$ and $\vect{y}$ into two parts: a deterministic equation that exactly matches the gating variable equation~\eqref{eq:x_HH}, and a fluctuation equation for the noise terms. To accomplish this we define new variables $\bar{\vect{x}}$ and $\hat{\vect{x}}$, which evolve via:
\begin{align} 
d \bar{\vect{x}} &= A_\K(V) \bar{\vect{x}} dt \label{eq:FL12D_K_Deterministic}\\
d \hat{\vect{x}} &= A_\K(V) \hat{\vect{x}} dt + S_\K(V, \bar{\vect{x}} + \hat{\vect{x}}) d \vect{W}(t)  \label{eq:FL12D_K_Stochastic},
\end{align}
with initial conditions $\bar{\vect{x}}(0) = \vect{x}(0)$ and  $\hat{\vect{x}}(0) = 0$. The sum  $\bar{\vect{x}} +  \hat{\vect{x}}$ solves Eq.~\ref{eq:FL12D_K}, so this is an exact decomposition of $\vect{x}$ into a deterministic part $\bar{\vect{x}}$ and a fluctuation part  $\hat{\vect{x}}$.  We can also apply a similar decomposition to $\vect{y}$.  As discussed by a number of authors~\cite{Dayan2001, Keener2009, Goldwyn2011}, solutions to the deterministic equation (Eq.~\ref{eq:FL12D_K_Deterministic}) can be generated by appropriate combinations of $m$, $n$, $h$, the gating variables from the deterministic HH equations: $\bar{y}_o=m^3 h$ and  $\bar{x}_o = n^4$.  This leaves the fundamental structure of the HH equations intact. Eq.~\ref{eq:V_OpenVar} can be replaced by the modified HH voltage equation (Eq.~\ref{eq:V_HH_ConductanceNoise}), where the conductance noise terms $\xi_\K(V,t)$ and $\xi_\Na(V,t)$ are defined to be $\hat{x}_o(t)$ and $\hat{y}_o(t)$, respectively.  

In sum, the high-dimensional SDEs derived by Fox and Lu~\cite{Fox1994} do not modify the deterministic structure of the HH equations.  Instead, as shown in Eq.~\ref{eq:FL12D_K_Stochastic}, their sole purpose is to shape the fluctuations in the fractions of open channels.  An important strength of this method is that it yields a description of channel fluctuations that is equally valid outside of voltage clamp.  Furthermore, as shown in~\cite{Goldwyn2011}, the stationary statistics of open channels for this method match exactly those of the Markov chain model, and it accurately replicates spiking statistics for channel numbers as small as 600 \Nap and 180 \Kp channels (membrane area of 10 $\mu m^2$).

One complication in solving these systems of SDEs is the need to determine $S_\Na(V,\vect{y})$ and $S_\K(V,\vect{x})$, by computing matrix square roots in each time step.  In order to guarantee the existence of these matrix square roots, we replace the values $\vect{y}$ and $\vect{x}$ in the diffusion matrices with deterministic values obtained from the gating variables, or equivalently the solutions Eq.~\ref{eq:FL12D_K_Deterministic} for $\vect{x}$ and the corresponding equation for $\vect{y}$.

\section*{Comparing stochastic versions of the Hodgkin-Huxley equations:  simulations }

\begin{figure}[!ht]
\begin{center}
\includegraphics[width=4in]{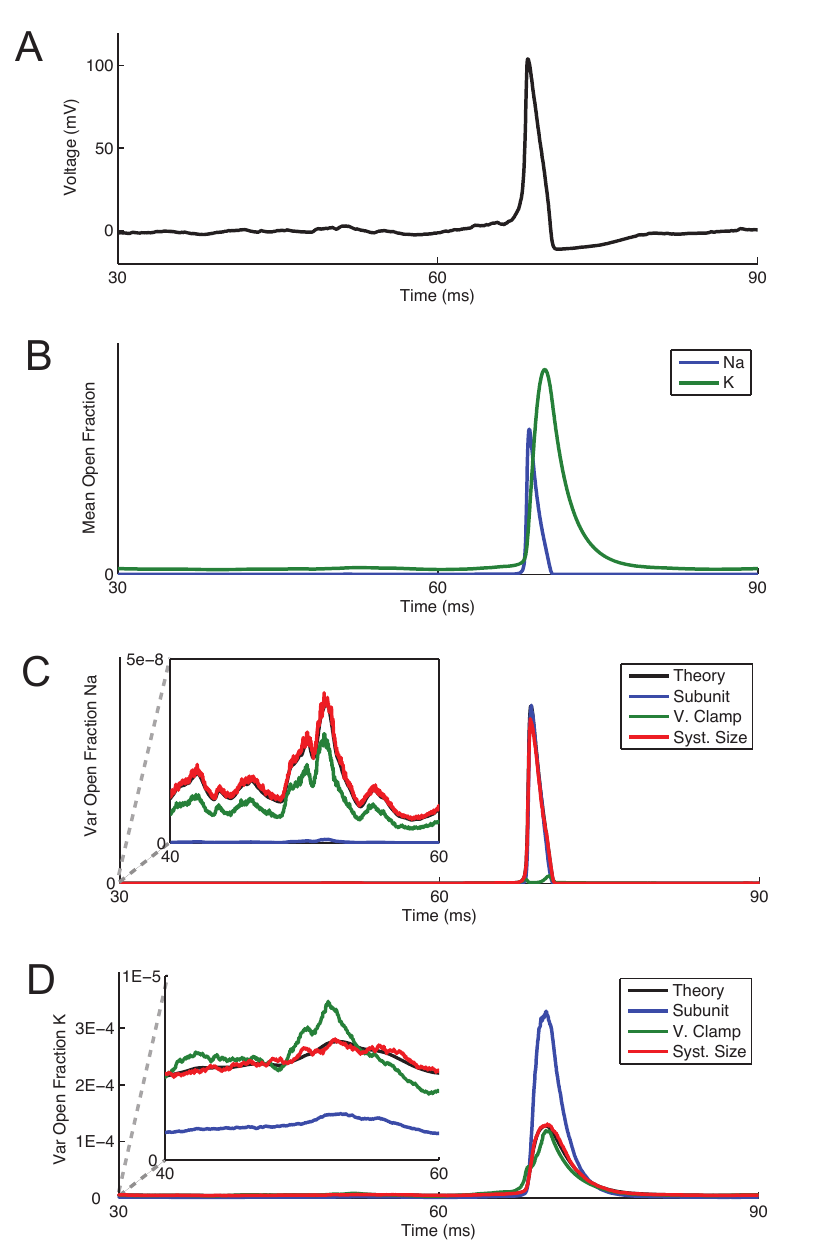} 
\end{center}
\caption{
{\bf Analysis of responses of channel noise models for a fixed voltage trajectory.}  
{\bf A:} Voltage trace obtained from the Markov chain model with no current input, 6000 \Nap channels and 1800 \Kp channels.  Dynamics are characterized by a prolonged subthreshold period followed by a spontaneous, channel-noise induced spike at $~70 ms$. 
{\bf B:} Means of fraction of open \Nap and \Kp channels for the given voltage trace in Panel A, as computed from Eqs.~\ref{eq:KeenerMeanNa} and~\ref{eq:KeenerMeanK}.  
{\bf C:} Variance in the fraction of open \Nap channels.  
{\bf D:} Variance in the fraction of open \Kp channels.  Left insets in Panels C and D show magnified views of the period preceding the spike. Right inset in Panel C shows magnified view during the spike.  For Panels C and D, exact variances (black) were computed from Eq.~\ref{eq:KeenerVarNa} and Eq.~\ref{eq:KeenerVarK} and all other variances were estimated from 5000 repeated simulations of the channel noise models.   
}
\label{fig:VTraceAndResponse}
\end{figure}

\begin{figure}[!ht]
\begin{center}
\includegraphics[width=4in]{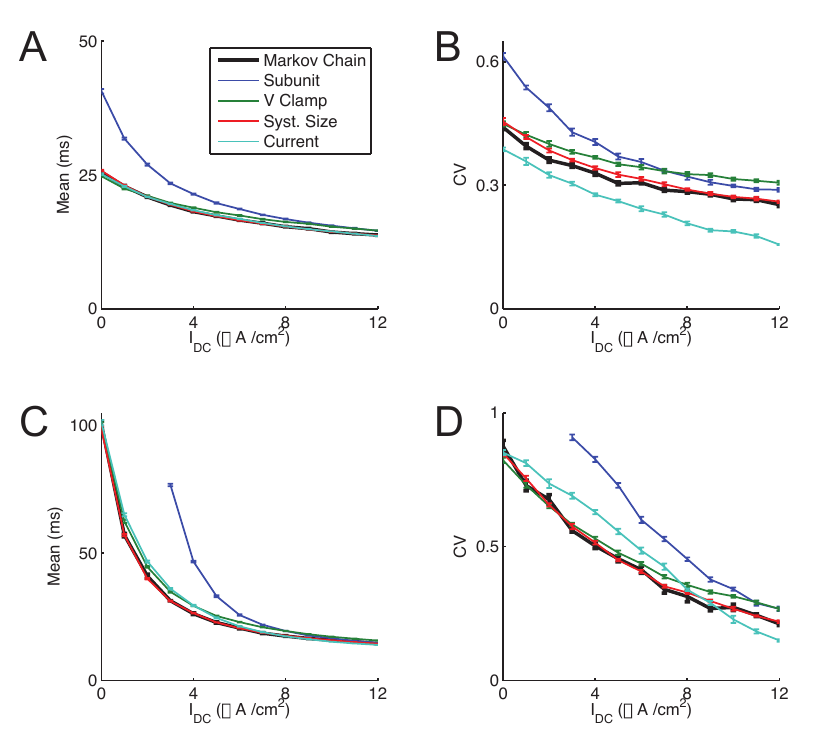}
\end{center}
\caption{
{\bf ISI statistics for DC input.}
{\bf A:} Mean of ISIs for a membrane area of $10 \mu m^2$ (600 \Nap and 180 \Kp channels).
{\bf B:} CV of ISIs for same membrane area as Panel A.  500 spikes were used to estimate the mean and variance, and error bars indicate standard error in the mean for 10 repeated measurements.
{\bf C:} Mean of ISIs for a membrane area of $100 \mu m^2$ (6000 \Nap and 1800 \Kp channels).
{\bf D:} CV of ISIs for same membrane area as Panel C.  500 spikes were used to estimate the mean and variance, and error bars indicate standard error in the mean for 10 repeated measurements for all models except the Markov chain model, for which only 4 repeated measurements were used.
 }
\label{fig:ISI_Noise0}
\end{figure}

\begin{figure}[!ht]
\begin{center}
\includegraphics[width=4in]{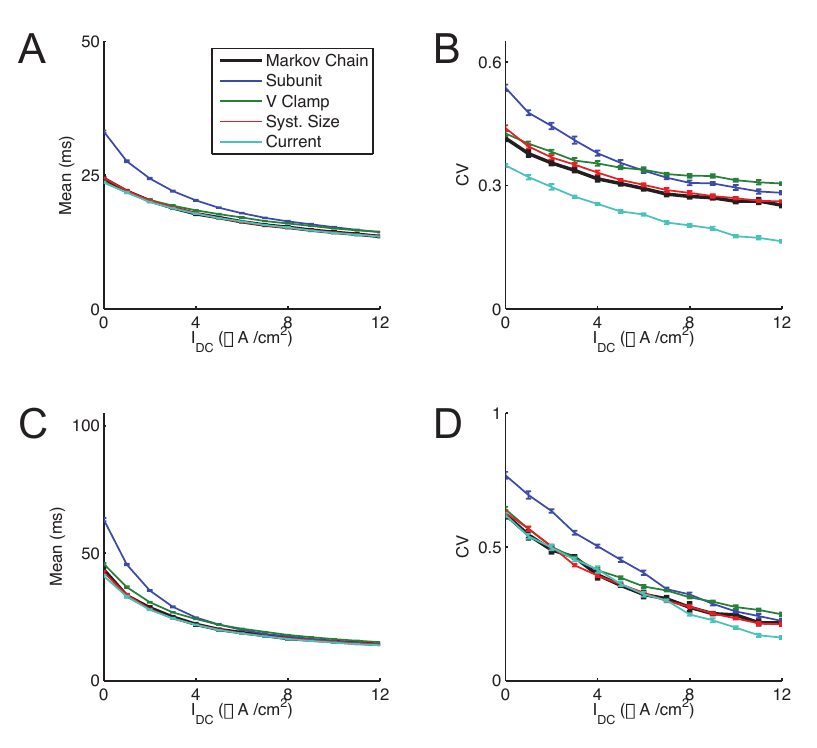}
\end{center}
\caption{
{\bf ISI statistics for noisy input.}
Same as Fig. \ref{fig:ISI_Noise0}, except that a noisy stimulus was used: $I = I_{DC} + 2 I_{noise}(t)$.
}
\label{fig:ISI_Noise2}
\end{figure}

How well do the simplified noise models match the ``gold standard'' Markov chain model of ion channel kinetics? 
Extensive comparisons between Markov chain and subunit noise models have been reported in prior studies~\cite{Mino2002, Zeng2004, Bruce2009, Sengupta2010, Goldwyn2011}.  Studies have also compared Markov chain models to a current noise model~\cite{Rowat2007}, voltage clamp conductance noise models~\cite{Goldwyn2011, Linaro2011}, and Fox and Lu's system size derived conductance model~\cite{Goldwyn2011, Orio2011}.  An exhaustive numerical investigation of these approaches is beyond the scope of this review, but in Figs.~\ref{fig:VTraceAndResponse}-\ref{fig:ISI_Noise2} we show simulation results that illustrate key differences among these approaches.  All simulations use standard parameter values for the HH equations \cite{Hodgkin1952}.  The voltage clamp conductance noise model is defined as in~\cite{Linaro2011}.  In all simulations, we used the Euler-Maruyama method with $0.01 ms$ time step for solving the relevant differential equations~\cite{Higham2001} and a Gillespie-type algorithm to implement the ion channel kinetics in the Markov chain~\cite{Gillespie1977, Chow1996}.  To generate Gaussian pseudorandom numbers, we produced uniform pseudorandom numbers with the Mersenne Twister algorithm \cite{mt19937:FortranVer} and then transformed these using the Box-Muller method \cite{Press1988}.   Simulation code is available upon request, and is based on the work of~\cite{Linaro2011} and~\cite{Goldwyn2011}.  Both of these groups have made their code available on the ModelDB website~\cite{Hines2004}, accession numbers 127992 and 128502, respectively.   To complement this review, we supply user-friendly Matlab simulation code of these stochastic versions of the Hodgkin-Huxley equations on the ModelDB website (accession number 138950) and at our website http://www.amath.washington.edu/$\sim$etsb/tutorials.html.

We will first compare time-varying distributions of the fractions of open channels.  Intuitively, one would expect that the number of open channels (all of which assumed to be independent), should be binomially distributed.  For a predefined voltage trajectory, this is indeed the case, as has been proven by ~\cite{Keener2009}.  The time-varying distributions of the fractions of open \Nap and \Kp channels in a Markov chain model of ion channel kinetics approach an asymptotically stable,  voltage dependent binomial distribution with means and variances given by solutions to the deterministic subunit equations Eq.~\ref{eq:x_HH}:
\begin{align}
&\mbox{\E[Fraction Open \Nap channels] }= m^3 h \label{eq:KeenerMeanNa} \\
&\mbox{\E[Fraction Open \Kp channels] }= n^4 \label{eq:KeenerMeanK} \\
&\mbox{\Var[Fraction Open \Nap channels]}  = \frac{m^3 h(1-m^3 h)}{N_\Na}  \label{eq:KeenerVarNa} \\
&\mbox{Var[Fraction Open \Nap channels]} = \frac{n^4(1-n^4)}{N_\K}  \label{eq:KeenerVarK}
\end{align}
We can use this result to compare channel noise models outside of voltage clamp.  Fig.~\ref{fig:VTraceAndResponse}A shows a single voltage trace obtained from a Markov chain model with 6000 \Nap channels and 1800 \Kp channels (membrane area $100 \mu m^2$) with no applied current ($I=0$).  Using this sample path as an input to the channel noise models, we compare the statistics of the fractions of open channels for the different models.  Panel B shows the mean fractions of open \Nap and \Kp channels, as computed from Eqs.~\ref{eq:KeenerMeanNa} and  \ref{eq:KeenerMeanK}.  All channel noise models produced mean values that were in close agreement with these values, so we did not plot those results.

The results for the variance of the fractions of open channels, as shown in Fig.~\ref{fig:VTraceAndResponse}C and \ref{fig:VTraceAndResponse}D, tell a different story.  The variance in the fractions of open \Nap channels are computed from Eqs.~\ref{eq:KeenerVarNa} and shown in black in Panel C.  The variance is accurately captured by Fox and Lu's conductance noise model (red), but misestimated by the subunit noise model (blue) and voltage clamp conductance noise model (green).  Of particular note is the fact that the voltage clamp conductance noise model utterly fails to track the Markov chain variance during the spike (right inset of Panel C).  This illustrates the point, made earlier, that voltage clamp methods may not be appropriate in regimes when $V$ changes rapidly.  The subunit noise model underestimates the variance during the subthreshold period (left inset), and overestimates the variance during the spike at $\sim 70 ms$ (right inset).

Fig.~\ref{fig:VTraceAndResponse}D shows variances in the fraction of \Kp channels.  Again, Fox and Lu's conductance noise model is most consistent with the equilibrium binomial distribution result. 
The voltage clamp model provides a reasonably close approximation, but the subunit noise model alternately undervalues the variance prior to the spike (see inset), and overvalues the variance near the time of the spike.

We also simulated spike trains from the channel noise models, using a current input of the form
\begin{align}
I = I_{DC} + I_{noise} \xi_I(t)
\end{align}
where $\xi_I(t)$ is a Gaussian white noise process with zero mean $\E[\xi_I(t) \xi_I(t')] =\delta(t-t')$. In Fig.~\ref{fig:ISI_Noise0}, we show the mean and coefficients of variance (CV) of interspike intervals (ISIs) obtained from simulations of the Markov chain and SDE models in response to a constant current input ($I_{noise}=0 \mu A/cm^2$).  Similar simulation results have been reported in~\cite{Zeng2004, Rowat2007, Sengupta2010, Goldwyn2011}.  We present results for different amounts of constant current input (x-axis) and two different membrane areas.  The magnitude of fluctuations in the current noise model was chosen so that the mean insterspike interval of this model would match that of the Markov chain model:  $\xi_V(t) = 1.94 \eta(t)$ for a membrane area of $10 \mu m^2$ and  $\xi_V(t) = 4.3 \eta(t)$ for a membrane area of $100 \mu m^2$, where $\eta(t)$ is a Gaussian white noise process with mean zero and $\E[\eta(t)\eta(t')] = \delta(t-t')$.

In the left column, we see that all models, with the known exception of the subunit noise model (blue), accurately reproduce the mean ISIs of the Markov chain (black), although there are slight discrepancies apparent for the current noise (cyan) and voltage clamp (green) methods.  These discrepancies are even more visible when comparing the coefficient of variation of the ISIs (right column).   Fig.~\ref{fig:ISI_Noise2} presents the results of similar simulations, the only difference being the addition of Gaussian white noise fluctuations to the applied input ($I_{noise}=2 \mu A/  cm^2$).  The ISI statistics for the subunit noise model remain the most different from the Markov chain.   For all conditions tested, it is clear that the Fox and Lu's conductance noise model (red) generates ISI statistics that are most similar to the Markov chain model.

\section*{Discussion}

We stand at a promising moment for the study of channel noise in conductance-based models.  In recent years, due to a spate of simulation studies drawing attention to discrepancies between subunit noise models and Markov chain ion channel models~\cite{Mino2002, Zeng2004, Bruce2009, Sengupta2010, Goldwyn2011, Linaro2011}, there has been a growing sense of pessimism regarding whether SDEs could prove an effective framework for modeling the stochastic activity of populations of ion channels \cite[e.g.]{Faisal2010}.   However, thanks to the development of novel approximation methods~\cite{Goldwyn2011, Linaro2011} and the rediscovery, analysis, and testing of past efforts~\cite{Fox1994, Fox1997}, new life has been breathed into the SDE approach.  The validity of SDE versions of the HH equations is now more clearly justified, and the door is open for these models to generate insight into how channel effects spike timing, reliability, propagation, and other aspects of neural dynamics.

A central theme of this review is that the addition of fluctuations in conductance terms, or equivalently in the fractions of open channels, should be the preferred way for including channel noise in the HH equations.  This approach, which we have termed \emph{conductance noise}, generates models that preserve the mathematical structure of the HH equations and that accurately approximate Markov chain models.  In the case of the high-dimensional SDE model derived by Fox and Lu in~\cite{Fox1994}, this was not obvious at first glance, and may be one reason why this aspect of their work has been overlooked.  Through a brief calculation, however, we elucidated the connection between this model and the HH equations by showing how the high-dimensional SDEs can be decomposed into a deterministic part identical to the classical HH equations and a fluctuation part representing channel noise.  

Although SDE models for channel noise are generally validated by making comparisons to Markov chain versions of the HH model, there is no guarantee that the Markov chain framework will remain the ``gold standard.''
Indeed, critiques of the Markov chain approach have been articulated~\cite[cf.]{Jones2006} and alternative mathematical models have been proposed~\cite[e.g.]{Liebovitch2001}.   With this in mind, it is useful to draw a distinction between ``derived models''  and ``empirical models.''  The subunit and conductance noise models introduced by Fox and Lu~\cite{Fox1994, Fox1997}
 are in the former category.  They are constructed with explicit reference to the conformational states of ion channels and their subunits, as defined by a Markov chain model of ion channel kinetics.  In contrast, the current noise model
 and the voltage clamp conductance noise models
 can be thought of as ``empirical'' since they can be constructed from observable quantities.  In our simulations, for instance, we used spontaneous firing rate to set the current noise level and the stationary statistics of open channels in the Markov channel model to define the noise processes in the voltage clamp conductance models.  In principle, empirical measurements of conductance fluctuations in voltage clamp, without reference to a Markov chain model, could be used to construct conductance fluctuations.  Empirical models that can be fit to, or validated against, quantities that are readily available from electrophysiological data are an attractive direction for future research, as they may inspire new methods for incorporating channel noise in conductance-based models.

The effects of channel noise have been a subject of intense interest in computational neuroscience and related fields in computational biology.  The stochastic approaches reviewed in this paper represent an important extension of the conductance-based model framework introduced by Hodgkin and Huxley~\cite{Hodgkin1952}.  Due to decades of analysis of the HH equations and an abundance of theoretical tools~\cite{Freidlin1998} and numerical methods \cite[e.g.]{Alzubaidi2010} for studying SDE models, we believe that appropriate methods for adding noise processes to the HH equations and their cousins throughout electrophysiology will play an important role in the future of computational biology.


\section*{Acknowledgments}
We thank Jay Rubinstein for drawing our interest to this problem.   In addition, we are grateful to Hong Qian, Mike Famulare, and members of the Shea-Brown research group for many helpful discussions and comments on the mansucript.  Funding support was provided by the National Institute on Deafness and Other Communication Disorders  (F31 DC010306---J.H.G.) and the Burroughs--Wellcome Fund (Career Award at the Scientific Interface---E.S.-B.).

\bibliography{bib}

\end{document}